\pdfoutput=1
\documentclass[acmsmall,nonacm]{acmart}

\usepackage{graphicx}
\usepackage{subcaption}
\usepackage{makecell}

\usepackage{listings}
\usepackage{upquote}

\usepackage{float}
\usepackage{tikz}
\usepackage{xcolor}
\usetikzlibrary{arrows.meta}

\definecolor{lcInk}{HTML}{353531}
\definecolor{lcJava}{HTML}{FF9505}
\definecolor{lcProbe}{HTML}{EC4E20}
\definecolor{lcKern}{HTML}{3A7D44}

\newcommand{\lcrule}{0.9pt}
\newcommand{\lcthin}{0.7pt}

\lstdefinestyle{numbered}{
  numbers=left,
  numberstyle=\tiny\color[gray]{0.5},
  numbersep=6pt,
  xleftmargin=1.4em,
}

\lstdefinelanguage{dtrace}{
  sensitive=true,
  morekeywords={provider,probe,self,
                int,uintptr_t,long,char,string,void},
  morecomment=[l]{//},
  morecomment=[s]{/*}{*/},
}

\lstdefinelanguage{bpftrace}{
  sensitive=true,
  alsoletter={@},
  morekeywords={BEGIN,END,usdt,uprobe,tracepoint,interval,
                if,else,delete,exit,printf,print,str,count,
                pid,tid,comm,nsecs,args,arg0,arg1,arg2,arg3},
  morecomment=[l]{//},
  morestring=[b]",
}

\AtBeginDocument{%
  }

\begin{document}

\title{Towards Runtime-Aware Observability for Java Virtual Threads}

\author{Wenjie Ding}
\email{wenjie.ding77@gmail.com}
\affiliation{%
  \institution{School of Computing, Newcastle University}
  \city{Newcastle upon Tyne}
  \country{United Kingdom}
}

\begin{abstract}
OS-level observability tools join application requests to kernel events through the OS thread ID, assuming that a unit of work occupies one OS thread until it completes. Java virtual threads break this assumption: many virtual threads are multiplexed onto a few carrier threads, so a stable thread ID hides a changing unit of work, and the join fails silently. Existing virtual-thread-aware channels republish context or identity on every mount and unmount, and the JVMTI-based one pays this cost for the lifetime of the JVM, whether or not an observer is attached. We bind a 64-byte off-heap trace buffer to each virtual thread, at a fixed address for its lifetime, and publish the address to eBPF consumers through four USDT probes in a patched OpenJDK; a re-implemented JVMTI channel over the same buffer serves as a controlled comparison. The four-probe design attributed all 768 test markers correctly on both the content and the identity plane, while the naive two-probe design silently misattributes trace context. With no consumer attached, publication costs at most about 33\,ns per transition; the remaining dormant cost is buffer allocation, about $0.37\,\mu$s per virtual thread, which is negligible on transition-heavy workloads but nearly doubles the cost of a creation-only microbenchmark. The JVMTI channel costs about 20\% on transition-heavy workloads and about $1.8\,\mu$s per virtual thread on the creation-only one. We also show that, of the two update strategies OTEP 4947 permits, only in-place rewriting guarded by the record's valid flag is sound for readers that cache the record's address, as lifecycle-event readers do.
\end{abstract}

\maketitle

\section{Introduction}

\subsection{Motivation}
\label{Motivation}
Observability data comes from two layers that see different things. The application knows which request it is serving; the operating system knows what that work costs in CPU time, blocking and I/O. Neither can answer on its own the question operators actually ask: which request caused this CPU consumption, or this blocking? Answering it requires a \emph{join}: attaching application-level request information to events observed at the OS level.

In practice, the join key is the OS thread ID. Profilers aggregate samples by thread ID, eBPF programs use it as their map key, and zero-code tracers such as DeepFlow, Grafana Beyla and vPath correlate spans by it~\cite{crosstrace}. The key is correct only if each unit of work occupies one OS thread exclusively until it completes, an assumption we call \emph{one-execution-per-thread}. No specification states it, yet tools rely on it widely, so when it breaks, the join fails silently. OpenTelemetry's eBPF instrumentation shows what this looks like in practice: with virtual threads enabled, 63\% of downstream spans lost their parent request, and 19\% of log lines carried another request's trace ID~\cite{obi2242,obi2284}.

Java virtual threads~\cite{jep444} break this assumption by design. They let a server keep the thread-per-request programming model while serving far more concurrent requests than it has OS threads, by scheduling many virtual threads onto a few carrier threads ($M:N$ scheduling). To the OS, a carrier's thread ID stays the same while the unit of work it runs keeps changing. This is not a defect but the source of the performance benefit, since switching between virtual threads involves no kernel transition; making virtual threads visible to the OS would defeat their purpose.

Existing and emerging standards let a runtime publish its trace context to outside readers, but only at process or OS-thread granularity~\cite{otep4719,otep4947}. Virtual-thread-level attribution has nevertheless been achieved: Elastic uses a JVMTI agent~\cite{elasticotel}, and OpenTelemetry's eBPF instrumentation has since tracked virtual-thread mounts through bytecode instrumentation~\cite{obi2261}. Both pay on the mount/unmount path, so the question is no longer whether attribution is possible, but what it costs. We ask three questions:
\begin{description}
  \item[Q1] Can the mapping from carrier thread to virtual thread be reconstructed at the OS level without the JVMTI capability?
  \item[Q2] Can this be free while no consumer is attached?
  \item[Q3] What determines the overhead of each publication channel?
\end{description}
To answer them, we modified OpenJDK to publish a per-virtual-thread trace buffer through virtual-thread lifecycle events consumed by an eBPF program, and re-implemented the existing JVMTI-based channel over the same buffer protocol as a controlled comparison.

\subsection{Contributions}
\begin{itemize}
  \item \textbf{A characterisation of the correlation problem under $M:N$ scheduling.}
  We apply soundness and completeness to virtual-thread attribution and evaluate it on two separate planes (Section~\ref{Correctness Definitions}): the content plane (is the trace context correct?) and the identity plane (is the event attributed to the right virtual thread?). A trace context can be correct while attributed to the wrong virtual thread, and a single-plane check accepts such cases.

  \item \textbf{A protocol-level finding about OTEP 4947.}
  OTEP 4947 lets writers update the published context in either of two ways: by swapping the record pointer, or by rewriting a fixed record in place, guarded by the record's valid flag. Both are safe under the reader model the OTEP implicitly assumes, in which readers re-resolve the record's address on every read. A virtual thread has no OS-resolvable identity, so a reader must either rely on the runtime republishing a carrier-side pointer on every transition, or cache the address it learns from lifecycle events, as ours does. We make this assumption explicit and show that, for caching readers, only in-place rewriting is sound: a pointer swap leaves the cached address pointing at a stale or released record (Section~\ref{Trace Buffer Design}). Our experiments observe exactly this failure: a reader holding a stale address returns a well-formed context from a released buffer (Section~\ref{Correctness}).

  \item \textbf{A mechanism and a decomposition of its cost.}
  We bind an off-heap buffer to the lifecycle of each virtual thread in OpenJDK and publish its address through four USDT probes. We show that the naive two-probe (freeze/thaw) reconstruction is not merely incomplete but unsound: it silently misattributes trace context. A second channel over the same buffer, built on JVMTI, isolates the capability tax. While dormant, our publication path costs at most about 33\,ns per transition, whereas the JVMTI channel costs about 90--105\,ns per transition (around 20\% on transition-heavy workloads) and about $4.6\times$ our channel's per-thread cost on a creation-only workload (Section~\ref{Evaluation}).
\end{itemize}

The rest of this paper is organised as follows. Section~\ref{Background} introduces virtual threads, OpenTelemetry trace context, USDT/eBPF and JVMTI, and Section~\ref{Related Work} reviews existing approaches to exposing trace context. Section~\ref{Problem Characterisation} characterises the correlation problem and defines the correctness criteria. Sections~\ref{Design} and~\ref{Implementation} present the design and implementation of our mechanism, and Section~\ref{Evaluation} evaluates its correctness and cost against the JVMTI channel. Section~\ref{Discussion} discusses trade-offs and limitations, and Section~\ref{Conclusion} concludes.

\section{Background}
\label{Background}
\subsection{Virtual Threads}
\label{Virtual Threads}
Virtual threads were introduced in Java 21~\cite{jep444} to preserve the thread-per-request style of server applications while decoupling application concurrency from the number of OS threads. The JVM schedules many virtual threads ($M$) onto a small pool of carrier threads, which are ordinary platform threads ($N$). Attaching a virtual thread to a carrier is called \emph{mounting}; detaching it, \emph{unmounting}.

When a virtual thread parks, for example on a blocking operation, the runtime copies its stack frames to the heap (\emph{freeze}) and unmounts it; when it is resumed, the frames are copied back onto a carrier (\emph{thaw}). Freeze and thaw are implemented once, in the HotSpot C++ runtime, and all three execution engines (the template interpreter and the C1 and C2 compilers) converge on them. Two transitions bypass them: the first mount of a virtual thread performs no thaw, since nothing has been frozen yet, and the final unmount, when the task completes, performs no freeze.

\subsection{OpenTelemetry and Trace Context}
\label{OpenTelemetry Trace Context}
A trace is the collection of all work done to serve a single request; each unit of work within a trace is a span~\cite{36356}. Request-level tracing has a long line of systems work, from X-Trace and Dapper to Pivot Tracing and Canopy~\cite{xtrace,36356,pivottracing,canopy}; this work addresses one link in that chain, the join between OS-level events and the request executing on a thread. OpenTelemetry, a vendor-neutral observability framework that evolves through OpenTelemetry Enhancement Proposals (OTEPs), follows the W3C Trace Context specification~\cite{w3ctracecontext}, whose core fields are the trace ID and the span ID. The OpenTelemetry SDK keeps a \emph{current context}, including the current span, for each thread, and all changes to it go through the \texttt{ContextStorage} interface.

OTEP 4947~\cite{otep4947}, merged in July 2026, specifies how an SDK publishes the active trace context to external readers. The SDK maintains a record for each thread, holding the trace ID, the span ID and a valid flag, and publishes a pointer to it in an ELF (Executable and Linkable Format) TLS (Thread-Local Storage) variable, so that an outside reader, such as the OpenTelemetry eBPF profiler~\cite{otelebpfprofiler}, can locate a thread's record through a known symbol. Writers may update the context either by swapping the TLS pointer to a different record, or by rewriting a fixed record in place while its valid flag is cleared. The protocol requires readers to read only while the target thread is stopped or interrupted, and implicitly assumes that they resolve the address afresh on every read.

\subsection{USDT, Tracepoints and eBPF}
A User Statically-Defined Tracing (USDT) probe is an instrumentation point compiled into a program, described by an ELF note that records its name, location and arguments~\cite{dtrace}. While no tracer is attached, the probe site is a \texttt{NOP}; attaching a tracer replaces it with a trap into the kernel. A kernel \emph{tracepoint} is the kernel-side counterpart: it is defined independently of the observed application and fires in the context of the calling task, so it sees the task's thread ID but no application state~\cite[\S2.9]{BPFPerformanceTools}. eBPF programs can be attached to both, USDT probes through uprobes, and keep state across events in key--value maps. An eBPF handler runs synchronously in the context of the task that triggered it: until the handler returns, that task executes no user-space code~\cite[Ch.~2]{BPFPerformanceTools}.

\subsection{JVMTI and Virtual Threads}
\label{JVMTI and Virtual Threads}
The JVM Tool Interface (JVMTI)~\cite{jvmti21} lets a native agent, running inside the JVM process, request capabilities and receive callbacks for registered events; Elastic's implementation is built on it (Section~\ref{The Elastic JVMTI Implementation}). For virtual threads, JVMTI support has a cost that does not depend on what the agent does. In HotSpot (JDK 21), as soon as any JVMTI environment is created, whether by an agent loaded at start-up or by one attached later, every mount and unmount is routed through a notifying transition path, and in production this is never switched off again; the HotSpot source itself notes that this has a performance impact.\footnote{\texttt{JvmtiExport::get\_jvmti\_interface} in \texttt{src/\-hotspot/\-share/\-prims/\-jvmtiExport.cpp} (JDK 21): ``There is a performance impact when VTMS transitions are enabled.'' The only function that disables them is a WhiteBox testing hook, documented as unsafe while virtual threads are running.} The cost is therefore paid on every transition, whether or not an observer is attached. We call it the \emph{capability tax}.

\section{Related Work}
\label{Related Work}

\subsection{Observability from Inside the Runtime}
\label{Observability from Inside the Runtime}
Some runtimes already attribute work to their user-level threads, but with the consumer inside the runtime. JDK Flight Recorder (JFR)~\cite{jep328} attributes its own events, such as I/O, allocation, and virtual-thread events like \texttt{jdk.VirtualThreadStart} and \texttt{jdk.VirtualThreadPinned}~\cite{jep444}, directly to virtual threads, without needing the carrier. Its producer and consumer both live inside the JVM, and its consumer path is asynchronous. Our consumer, by contrast, sits outside the process, must attribute each event at the moment it occurs, and needs the carrier thread ID as the bridge to OS-level events (Section~\ref{The Correlation Problem under M:N Scheduling}).

Go's goroutines~\cite{go} are the closest analogue to virtual threads. Go maintains goroutine profiler labels that profiling infrastructure can associate with execution~\cite{gopprof}, and an external reader can start from the OS thread and walk the runtime's own structures to the running goroutine's labels, which is how eBPF profilers such as Parca attribute trace context to goroutines~\cite{parca}. Java offers no such path: starting from a carrier thread, there is no externally resolvable location for the mounted virtual thread's context. The closest candidate, the \texttt{VirtualThread} object itself, can be moved by the garbage collector, so an external reader cannot cache its address. Our mechanism provides exactly this missing piece: an off-heap address that is stable for the virtual thread's lifetime.

\subsection{The Elastic JVMTI Implementation}
\label{The Elastic JVMTI Implementation}
The Elastic Distribution of OpenTelemetry Java~\cite{elasticotel} attributes trace context to virtual threads through a JVMTI agent that registers the \texttt{VirtualThreadMount}/\texttt{Unmount} extension events. Every thread, platform or virtual, has an off-heap context buffer; on each mount, the agent writes the virtual thread's buffer address into a TLS slot of the carrier thread, and on each unmount it clears the slot. This is a pointer update rather than a copy of the record, and it preserves the resolve-at-read model: carriers are OS threads, so an OS-side reader resolves their TLS exactly as it would for platform threads.

The price is that visibility must be republished on every mount and unmount, and that the agent pays the capability tax (Section~\ref{JVMTI and Virtual Threads}) whether or not an observer is attached. Both costs fall on the transition path that virtual threads are designed to make cheap: the more often a workload transitions, the more it benefits from virtual threads, and the more it pays.

\subsection{OpenTelemetry eBPF Instrumentation}
\label{OBI}
Concurrently with this work, OpenTelemetry eBPF Instrumentation (OBI) met the same problem from the reader's side~\cite{obi2242}. Its Java agent keyed in-process trace context by kernel thread ID; with virtual threads enabled, the carrier changed between a handler's entry and its outgoing call in 91.5\% of requests, and 63.3\% of downstream spans lost their parent. The fix~\cite{obi2261} uses bytecode instrumentation (ByteBuddy) at the exit of \texttt{VirtualThread.mount()} and \texttt{unmount()}: on every transition, the agent reports the mounted virtual thread's ID to eBPF through an \texttt{ioctl}, and eBPF keeps a carrier-to-virtual-thread map. Trace-context publication through the OTEP map is nevertheless suppressed while a virtual thread is mounted: keeping it would have left 48\% of log lines with a wrong trace ID, and the project chose no data over wrong data~\cite{obi2284}. Restoring it remains an open issue, and the cost of refreshing context on every transition has not been measured.

OBI is therefore a third publication channel, alongside JVMTI and our USDT probes. Like the JVMTI channel, it pays on every mount and unmount, here with a system call, and it reports correctness but no overhead measurements. Its experience also confirms, on a production tracer, the failure mode our correctness criteria are designed to catch: wrong attribution that looks like valid data (Section~\ref{Correctness Definitions}).

In short, virtual-thread-level attribution is already achievable, but both existing channels pay on every mount and unmount, and the JVMTI channel pays its tax for the lifetime of the JVM, whether or not an observer is attached. This work asks whether that cost can instead be made conditional on an observer being present.

\subsection{Trace Context Exposure: A Three-Level Framework}
We compare approaches to exposing trace context along two axes (Table~\ref{tab:three-level-framework}). The first is the identity the context is bound to: a process, an OS thread, or a logical unit of work. The second is how a reader obtains the context's address: by resolving it on every read (\emph{resolve-at-read}, typically through a symbol lookup), or by receiving it once from the runtime and caching it (\emph{publish-once-and-cache}). The second axis is rarely discussed, because for platform threads the two models make no practical difference; for virtual threads, they do (Section~\ref{Trace Buffer Design}).

OTEP 4719~\cite{otep4719} binds context to a process, publishing it in a named memory mapping that readers locate through \texttt{/proc/<pid>/maps}; it can tell which service an execution belongs to, but not which request. OTEP 4947 binds context to an OS thread: readers locate each thread's record in ELF TLS through a known symbol. It does not require runtimes for which OS-thread-local context is the wrong execution-context model to implement this mechanism; such runtime-specific discovery mechanisms are left out of scope, but the OTEP recommends that they reuse its record format. A virtual thread has no OS-resolvable identity, so resolve-at-read can reach it only through its carrier, and the runtime must then republish on every mount and unmount: a buffer pointer in Elastic's case, the virtual-thread ID in OBI's. Our mechanism is the runtime-specific alternative the OTEP leaves room for: it reuses the record format and replaces TLS lookup with lifecycle events and cached addresses. The three levels are not a ranking: each serves a different reader model, and resolve-at-read is the right choice for the readers OTEP 4719 and 4947 target.

\begin{table}[t]
  \caption{Three levels of trace-context exposure.}
  \centering
  \small
  \setlength{\tabcolsep}{3pt}
  \label{tab:three-level-framework}
  \begin{tabular}{@{}llll@{}}
    \toprule
    Level & Bound to & Address discovery & Attributable unit \\
    \midrule
    Process (OTEP 4719) & process & named mapping, \texttt{/proc} & service \\
    Thread (OTEP 4947)  & platform thread & symbol, resolve-at-read & OS thread \\
    Virtual thread (this work) & vthread lifetime & lifecycle event, cached & unit of work \\
    \bottomrule
  \end{tabular}
\end{table}

\section{Problem Characterisation}
\label{Problem Characterisation}
\subsection{The Correlation Problem under M:N Scheduling}
\label{The Correlation Problem under M:N Scheduling}
At the OS level, an event carries exactly one execution identity: the thread ID (\texttt{tid}). Attributing it to a unit of work takes two steps,
\[
  \mathit{event}(\mathit{tid}, t) \;\rightarrow\; \mathit{vthread} \;\rightarrow\; \mathit{trace\ context},
\]
first determining which virtual thread was running on the carrier at time $t$, then reading that virtual thread's trace context at that moment. Under one-execution-per-thread, the first step is trivial: the \texttt{tid} itself is the answer. Under $M:N$ scheduling, the answer changes over time, and virtual-thread scheduling produces no event the OS can see, such as a system call, a context switch or a scheduler tracepoint. A consumer can therefore hold an expired mapping without any way of knowing it is stale. The two steps also depend on different components: the first on the runtime, which knows what is mounted, and the second on the observability framework, which maintains the trace context. Each can fail independently, which motivates evaluating them separately.

\subsection{Correctness Definitions}
\label{Correctness Definitions}
\paragraph{C1: Attribution}
Every OS-level event that user code causes on a carrier thread must be attributed to the virtual thread that was running at that moment, with the trace context that was current in Java at that moment (the ground truth).

\paragraph{C2: Separate planes}
Trace-context correctness (the \emph{content plane}: trace ID and span ID) and virtual-thread attribution (the \emph{identity plane}) must be evaluated separately. They are maintained by different components: the observability framework writes the trace context, while the observed runtime provides the virtual-thread identity. Checking the content alone cannot distinguish a mechanism that works from one that fails but happens to return the right context.

\paragraph{C3: Soundness and completeness}
A mechanism is \emph{complete} if every such event receives an attribution, and \emph{sound} if every attribution it makes is correct. An incomplete mechanism fails loudly: the consumer sees that an event is unattributed. An unsound mechanism fails silently: a misattributed record carries a trace ID and span ID that are valid in both format and content, and nothing in the record reveals that it belongs to another virtual thread. For observability, unsoundness is therefore more dangerous than incompleteness. Such failures are not hypothetical: a stale mapping can still resolve to a well-formed record (Section~\ref{Correctness}).

\section{Design}
\label{Design}
OTEP 4947's symbol-based TLS lookup requires a thread identity the kernel can see, which virtual threads deliberately lack (Section~\ref{Virtual Threads}). This constraint narrows the design space to two representative designs: JVMTI event callbacks, and USDT lifecycle probes consumed by an eBPF reader.

\subsection{The Root Design Question: Where the Buffer Address Lives}
\label{The Root Design Question: Where the Buffer Address Lives}
The first decision is whom the per-thread trace buffer belongs to: the carrier thread or the virtual thread.
\begin{itemize}
    \item \textbf{Carrier (Elastic-style).} An OS-side reader resolves the carrier's TLS as usual, but the buffer address must be republished whenever a virtual thread mounts, so the total publication work grows linearly with the number of mount/unmount transitions, on top of the JVMTI capability tax.
    \item \textbf{Virtual thread.} The off-heap buffer lives exactly as long as its virtual thread. The consumer then only needs to know which virtual thread is running on which carrier, which can be derived from existing virtual-thread lifecycle events.
\end{itemize}
We choose the second option. Its cost is that OS-side consumers must become event-driven and adopt the publish-once-and-cache model.

\subsection{Trace Buffer Design}
\label{Trace Buffer Design}
OTEP 4947 implicitly assumes that each sampling operation locates the record through the thread's TLS pointer and reads it immediately; it does not consider readers that cache the record's address. The protocol's correctness rests on two preconditions: the read happens only while the target thread is stopped or interrupted, which the OTEP requires explicitly, and the address is resolved afresh on every read, which it assumes implicitly. For virtual threads, keeping resolve-at-read means republishing a carrier-visible TLS pointer on every transition, which adds a cost to every mount and unmount (Section~\ref{Two channel comparison}). Our design instead publishes the buffer address through USDT probes, and readers cache it. Producer and consumer remain unaware of each other: the virtual thread does not know whether a reader exists, and the reader learns about the buffer only through lifecycle events.

The reader model can now be stated precisely: an eBPF consumer receives the buffer address once per virtual-thread lifecycle event and keeps it for an arbitrary duration, from one thaw to the next freeze, while the buffer's contents may be rewritten any number of times without re-resolution. Every design decision that follows is a consequence of this model.

OTEP 4947 allows two strategies for updating the record and leaves the choice to the writer: pointer swap and in-place mutation. A pointer swap replaces the pointer rather than the record it points to, so the publication is atomic, and a reader that resolves the address on every read never sees a stale record. A caching reader, however, keeps the old address, and every subsequent read silently returns the stale context; if the freed record is reallocated to another virtual thread, the reader even returns that thread's context. In-place mutation keeps the record's address stable for the virtual thread's entire lifetime, so a cached address remains valid for as long as the buffer exists. This is exactly what a caching reader needs, but it moves the consistency problem into the buffer: the reader may observe a record in the middle of a rewrite. The general rule is that any signal placed outside the reader's access path is invisible to a caching reader. The validity signal must therefore sit where the reader already looks, inside the record itself: the valid flag. Of the two strategies the OTEP permits, only in-place mutation is sound for caching readers.

Can a reader observe a torn record when a read overlaps a write? Three properties rule this out or make it detectable. First, the writer follows the OTEP 4947 in-place rewrite protocol, \texttt{valid=0} $\rightarrow$ release fence $\rightarrow$ payload $\rightarrow$ release fence $\rightarrow$ \texttt{valid=1}: the first fence orders the cleared flag before any payload write, and the second orders the complete payload before the flag is set, so the design keeps the correctness the OTEP guarantees for its own reader model. Second, both of our readers, the USDT handler and the system-call tracepoint handler, run synchronously in the context of the writer's own carrier task, not in another process or thread. While a handler runs, the task is in the kernel and cannot execute user-space writer code, so reads and writes are ordered by the task's own program order. Third, this same-task argument requires that the task not enter the kernel between \texttt{valid=0} and \texttt{valid=1}, which JVM machinery such as safepoints could violate. If it does, the reader sees the flag at 0 and reports the event as invalid: an honest, detectable failure in the sense of C3 (Section~\ref{Correctness Definitions}). The flag thus signals both a rewrite in progress and, while no context is attached, the absence of context.

The buffer layout (Table~\ref{tab:buffer-layout}) follows OTEP 4947 with one exception: our prototype was built against an earlier draft in which byte 25 was reserved, whereas the merged OTEP defines it as \texttt{trace-flags}, which our prototype does not populate. The fixed 64\,B size follows the OTEP's suggestion that a frugal writer keep the whole record within a 64-byte cache line. The trace ID and span ID are stored byte by byte in W3C (network) byte order.
\begin{table}[t]
  \centering
  \small
  \caption{Trace buffer layout (64\,B per virtual thread), following OTEP 4947~\cite{otep4947}.}
  \label{tab:buffer-layout}
  \begin{tabular}{@{}lll@{}}
    \toprule
    \textbf{Bytes} & \textbf{Field} & \textbf{Content} \\
    \midrule
    0--15  & \texttt{traceId}         & W3C trace ID, network byte order \\
    16--23 & \texttt{spanId}          & W3C span ID, network byte order \\
    24     & \texttt{valid}           & 1: record complete; 0: being rewritten or no context \\
    25     & \texttt{trace-flags}     & W3C trace flags (reserved in the draft we implemented; not populated) \\
    26--27 & \texttt{attrs-data-size} & length of \texttt{attrs-data} \\
    28--63 & \texttt{attrs-data}      & encoded attributes (not used in this work) \\
    \bottomrule
  \end{tabular}
\end{table}

\subsection{USDT Probe Placement: Constraints, Dead Ends, and Four-Probe Decomposition}
\label{Probe Placement: Constraints, Dead Ends, and Four-Probe Decomposition}
The semantically natural place for the probes would be the mount and unmount boundaries themselves. We place them deeper in the runtime instead, so the implementation location differs from the semantic one. The reason is that a USDT probe needs native code on the execution path, and the question becomes where such code already exists, and what it would cost to add it where it does not (Table~\ref{tab:vthread-transitions}).

\begin{table}
  \caption{Logical transitions and their USDT probes.}
  \centering
  \small
  \setlength{\tabcolsep}{3pt}
  \label{tab:vthread-transitions}
  \begin{tabular}{lll}
    \toprule
    \textbf{Logical Transition} & \textbf{Native Code on Path} & \textbf{USDT Probe} \\
    \midrule
    remount (2nd to N-th time) & thaw---C++ runtime, shared by all three engines & \texttt{vthread\_\_thaw} \\
    non-final unmount (park) & freeze---same as above & \texttt{vthread\_\_freeze} \\
    first mount & \texttt{enterSpecial} stub enters Java directly, no thaw & \texttt{vthread\_\_start} \\
    final unmount (termination) & continuation returns normally, no freeze & \texttt{vthread\_\_end} \\
    \bottomrule
  \end{tabular}
\end{table}

Remounts and non-final unmounts are cheap to cover: freeze and thaw are independent of the execution engine, so probes placed there fire regardless of JIT compilation state. The first mount and the final unmount, in contrast, run in JIT-compiled Java code with no native call site on the path; this is not a defect but precisely why virtual threads are cheap. Introducing a native site there has three options, each with a cost:
\begin{enumerate}
  \item \textbf{A JNI call on every mount}, which adds a native transition to every mount, so the total work grows with the number of mounts.
  \item \textbf{A new intrinsic}, which would have to be implemented in the template interpreter, C1 and C2: too invasive for a tracing feature.
  \item \textbf{Piggybacking on the existing JVMTI hooks.} \texttt{notifyJvmtiMount}/\texttt{Unmount} are intrinsic candidates guarded by a flag, and when the flag is off, C2 compiles them away. A probe placed in their native implementations would fire under the interpreter but become unreachable once the caller is JIT-compiled; turning the flag on brings back the capability tax.
\end{enumerate}
The way out is that the first mount and the final unmount each happen only once per virtual thread. A native call at these two points therefore costs $O(1)$ per virtual-thread lifetime, and we add the \texttt{start} and \texttt{end} probes there.

\subsection{Consumer Model and Event-Level Equivalence}
\label{Consumer Model and Event-Level Equivalence}
The consumer is event-driven: it keeps a map from carrier thread ID to the virtual thread currently mounted on it. The map's size is bounded by the size of the carrier pool.

\paragraph{Event-level equivalence}
All the consumer needs to know is which virtual thread runs on which carrier, that is, when virtual threads mount and unmount. The stock JVM does not expose these transitions, but they can be reconstructed precisely from the four probes:
\begin{itemize}
    \item Mount $= \{\texttt{start}\} \cup \{\texttt{thaw(kind=top)}\}$: \texttt{start} covers the first mount, and \texttt{kind=top} selects exactly one event per logical remount.
    \item Unmount $= \{\texttt{freeze(result=ok)}\} \cup \{\texttt{end}\}$: \texttt{result=ok} corresponds to a successful park, and \texttt{end} to termination.
    \item $\texttt{freeze(result=pinned\_*)} \not\to$ unmount: the virtual thread is pinned and stays on its carrier, so the mapping is kept.
    \item $\texttt{thaw(kind=return\_barrier)} \not\to$ any transition: it is an internal frame-reconstruction step within a single logical remount and must be ignored.
\end{itemize}

\paragraph{Insertion and deletion rules}
The consumer inserts a mapping only on \texttt{start} and \texttt{thaw(kind=top)}; return-barrier thaws neither insert nor delete. We observed return-barrier thaws even after a virtual thread had terminated, so inserting on every thaw would reintroduce a stale mapping and, with it, silent misattribution. On \texttt{end}, the consumer deletes the mapping unconditionally, since the semantics of \texttt{end} are exact. Two counters in the consumer check these rules at run time: they count \texttt{freeze} and \texttt{end} events that find no mapping to delete. In all six four-probe correctness runs (Section~\ref{Correctness}), both counters stayed at zero, and the 128 insertions in each run were matched by exactly 128 deletions.

\paragraph{Window boundaries}
The reconstructed mount window deviates slightly from the semantic one, in opposite but safe directions. For a remount, the window is slightly wider: \texttt{thaw} fires before the Java-side mount completes. At the two lifecycle endpoints, it is slightly narrower: \texttt{start} fires after the first mount has completed, and \texttt{end} fires before \texttt{run()} returns, while the final unmount happens afterwards. In both directions, the mismatch covers only JVM-internal execution, and user code runs strictly inside the reconstructed window. At the edges, an event is either attributed to the correct virtual thread or explicitly left unmapped, never attributed to another virtual thread.

\subsection{Soundness and Completeness Analysis}
\label{Soundness and Completeness Analysis}
Our first design used only the two \texttt{freeze}/\texttt{thaw} probes, and it achieves neither property of C3 (Section~\ref{Correctness Definitions}). Completeness is lost at the first mount, which fires no \texttt{thaw}, and soundness is lost at termination, which fires no \texttt{freeze} and therefore leaves a stale mapping behind. Section~\ref{Correctness} confirms both failures experimentally.

\section{Implementation}
\label{Implementation}
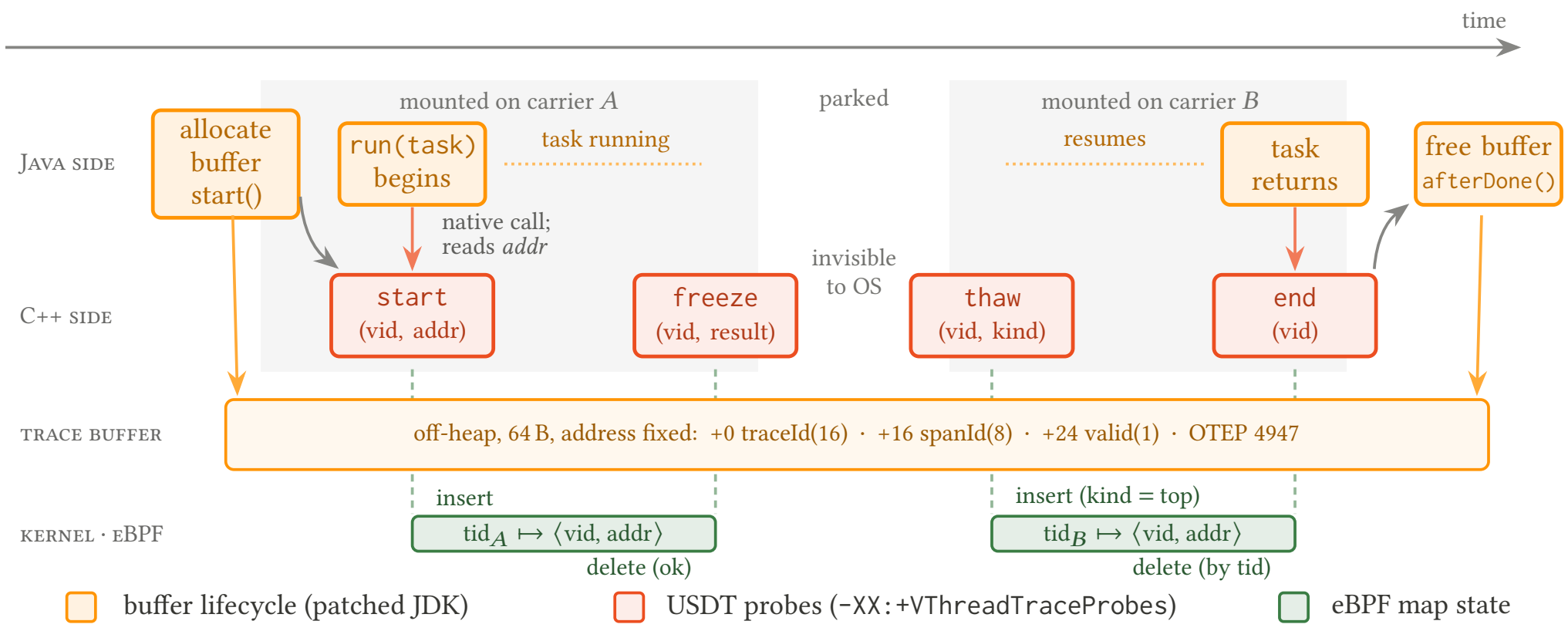
\begin{figure}[t]
  \begin{tikzpicture}[
  lcbox/.style={rounded corners=2pt, draw, line width=\lcrule, align=center,
    inner xsep=1pt, inner ysep=2pt, font=\scriptsize, minimum height=0.70cm},
  lcjava/.style={lcbox, text width=1.18cm, draw=lcJava, fill=lcJava!12,
    text=lcJava!70!black},
  lcprobe/.style={lcbox, text width=1.32cm, draw=lcProbe, fill=lcProbe!10,
    text=lcProbe!75!black},
  lcflow/.style={->, >=Stealth, line width=\lcthin},
  lcguide/.style={draw=lcKern!55, line width=\lcthin,
    dash pattern=on 1.8pt off 1.4pt},
  lctag/.style={font=\tiny\scshape, text=lcInk!75, anchor=west, align=left},
  lctiny/.style={font=\tiny, text=lcInk!80},
  lcphase/.style={font=\tiny, text=lcInk!70},
  lcswatch/.style={rounded corners=1pt, draw, line width=\lcthin,
    minimum width=7pt, minimum height=7pt, inner sep=0pt}
]

\fill[lcInk!5] (2.19,-1.78) rectangle (6.45,0.72);
\fill[lcInk!5] (8.10,-1.78) rectangle (11.50,0.72);
\node[lcphase] at (4.32,0.55) {mounted on carrier $A$};
\node[lcphase] at (7.28,0.55) {parked};
\node[lcphase] at (9.82,0.55) {mounted on carrier $B$};
\node[lcphase, align=center] at (7.28,-0.92) {invisible\\to OS};

\draw[->, >=Stealth, draw=lcInk!60, line width=\lcthin] (0.00,1.00) -- (13.0,1.00);
\node[font=\tiny, text=lcInk!70, anchor=south east] at (13.0,1.04) {time};

\draw[lcguide] (3.49,-1.76) -- (3.49,-3.00);
\draw[lcguide] (6.09,-1.76) -- (6.09,-3.00);
\draw[lcguide] (8.46,-1.76) -- (8.46,-3.00);
\draw[lcguide] (11.06,-1.76) -- (11.06,-3.00);

\node[lctag] at (0.00,0) {Java side};
\node[lcjava] at (1.89,0) {allocate buffer\\ start()};
\node[lcjava] at (3.49,0) {\texttt{run(task)}\\begins};
\node[lcjava] at (11.06,0) {task\\returns};
\node[lcjava] at (12.72,0) {free buffer\\{\tiny \texttt{afterDone()}}};

\draw[densely dotted, draw=lcJava!65, line width=\lcthin]
  (4.28,0) -- (5.97,0);
\draw[densely dotted, draw=lcJava!65, line width=\lcthin]
  (8.58,0) -- (10.29,0);
\node[font=\tiny, text=lcJava!70!black] at (5.15,0.20) {task running};
\node[font=\tiny, text=lcJava!70!black] at (9.43,0.20) {resumes};

\node[lctag] at (0.00,-1.30) {C++ side};
\node[lcprobe] at (3.49,-1.30) {\texttt{start}\\{\tiny (vid, addr)}};
\node[lcprobe] at (6.09,-1.30) {\texttt{freeze}\\{\tiny (vid, result)}};
\node[lcprobe] at (8.46,-1.30) {\texttt{thaw}\\{\tiny (vid, kind)}};
\node[lcprobe] at (11.06,-1.30) {\texttt{end}\\{\tiny (vid)}};

\draw[lcflow, draw=lcProbe!75] (3.49,-0.37) -- (3.49,-0.92);
\node[lctiny, anchor=west] at (3.62,-0.50) {native call;};
\node[lctiny, anchor=west] at (3.62,-0.70) {reads \emph{addr}};
\draw[lcflow, draw=lcProbe!75] (11.06,-0.37) -- (11.06,-0.92);

\draw[lcflow, draw=lcInk!60] (2.53,-0.28) to[bend right=24] (2.86,-0.90);
\draw[lcflow, draw=lcInk!60] (11.74,-0.90) to[bend left=24]  (12.04,-0.34);

\node[lctag] at (0.00,-2.32) {trace buffer};
\draw[draw=lcJava, fill=lcJava!7, line width=\lcthin, rounded corners=2pt]
  (1.89,-2.62) rectangle (12.72,-2.02);
\node[font=\tiny, text=lcJava!60!black] at (7.30,-2.32)
  {off-heap, 64\,B, address fixed:\ \ +0 traceId(16)
   \,$\cdot$\, +16 spanId(8) \,$\cdot$\, +24 valid(1)
   \,$\cdot$\, OTEP 4947};

\draw[lcflow, draw=lcJava!80] (1.95,-0.44) -- (1.99,-1.99);
\draw[lcflow, draw=lcJava!80] (12.66,-0.44) -- (12.62,-1.99);

\node[lctag] at (0.00,-3.17) {kernel\,$\cdot$\,eBPF};
\draw[draw=lcKern, fill=lcKern!13, line width=\lcthin, rounded corners=1.5pt]
  (3.49,-3.32) rectangle (6.09,-3.02);
\node[font=\tiny, text=lcKern!70!black] at (4.79,-3.17)
  {tid$_A \mapsto \langle$vid, addr$\rangle$};
\draw[draw=lcKern, fill=lcKern!13, line width=\lcthin, rounded corners=1.5pt]
  (8.46,-3.32) rectangle (11.06,-3.02);
\node[font=\tiny, text=lcKern!70!black] at (9.76,-3.17)
  {tid$_B \mapsto \langle$vid, addr$\rangle$};

\node[font=\tiny, text=lcKern!70!black, anchor=west] at (3.57,-2.86) {insert};
\node[font=\tiny, text=lcKern!70!black, anchor=east] at (6.01,-3.47) {delete (ok)};
\node[font=\tiny, text=lcKern!70!black, anchor=west] at (8.54,-2.86) {insert (kind $=$ top)};
\node[font=\tiny, text=lcKern!70!black, anchor=east] at (10.98,-3.47) {delete (by tid)};

\node[lcswatch, fill=lcJava!12, draw=lcJava] at (0.65,-3.8) {};
\node[right, font=\scriptsize, text=lcInk] at (0.89,-3.8)
  {buffer lifecycle (patched JDK)};
\node[lcswatch, fill=lcProbe!10, draw=lcProbe] at (5.35,-3.8) {};
\node[right, font=\scriptsize, text=lcInk] at (5.55,-3.8)
  {USDT probes (\texttt{-XX:+VThreadTraceProbes})};
\node[lcswatch, fill=lcKern!13, draw=lcKern] at (11.05,-3.8) {};
\node[right, font=\scriptsize, text=lcInk] at (11.25,-3.8)
  {eBPF map state};

  \end{tikzpicture}
  \caption{Lifecycle of a virtual thread's trace buffer: allocation and release on the Java side, the four USDT probes in HotSpot, and the eBPF map state they drive as the thread parks on carrier $A$ and remounts on carrier $B$.}
  \Description{A timeline with four lanes. The Java lane shows the buffer allocated in start() before the task runs and freed in afterDone() after it returns. The C++ lane shows the start, freeze, thaw and end probes. A trace-buffer lane spans the whole lifetime at a fixed address. The kernel lane shows the eBPF map entry from carrier A's thread ID to the virtual thread inserted at start and deleted at freeze, and an entry for carrier B inserted at thaw and deleted at end.}
  \label{fig:lifecycle}
\end{figure}

The implementation has three parts: the JDK allocates a buffer for each virtual thread and publishes its address (Sections~\ref{Buffer allocation}--\ref{Gating}), an eBPF consumer attributes OS-level events through it (Section~\ref{Consumers}), and an OpenTelemetry SDK plugin writes trace context into it (Section~\ref{SDK Integration}). Figure~\ref{fig:lifecycle} shows the buffer's lifecycle and the events that publish it.

\subsection{Buffer allocation}
\label{Buffer allocation}
Each virtual thread gets a 64\,B off-heap buffer, whose address is stored in a new field, \texttt{VirtualThread.\allowbreak traceBufferAddress}. We implemented two ways of allocating it: by the JVM itself, or by a JVMTI agent from outside the JVM. Section~\ref{Two channel comparison} compares them.

\subsubsection{JVM allocation}
\label{JVM allocate}
We modified OpenJDK\footnote{The modification adds 157 lines and removes 2, in 12 files across the Java and C++ sides of OpenJDK, on top of \texttt{jdk21u} commit \texttt{57b3005fc5}, branch \texttt{vthread-lifecycle-probes}~\cite{jdkpatch}.} so that it allocates the buffer itself. The allocation points follow one rule: the buffer's lifetime must strictly contain every interval during which a consumer holds its address, or the consumer may read freed memory. The buffer is therefore allocated in \texttt{VirtualThread.start()} and freed in \texttt{afterDone()}.

\subsubsection{JVMTI agent}
The JVMTI channel leaves the JVM's allocation path untouched: an agent allocates the buffer in its \texttt{VirtualThreadStart} callback, writes the address into the same \texttt{trace\allowbreak Buffer\allowbreak Address} field through JNI, and frees the buffer in \texttt{VirtualThreadEnd}; the JVM dispatches both callbacks from its existing \texttt{notifyJvmti\allowbreak Start}/\texttt{End} hooks. The agent is our own re-implementation of the Elastic-style design~\cite{elasticotel}, not Elastic's code: it requests only the \texttt{can\_\allowbreak support\_\allowbreak virtual\_\allowbreak threads} capability and, in its default mode, registers only the \texttt{VirtualThreadStart}/\texttt{End} events, besides \texttt{VMInit}. An optional \texttt{publish=jvmti} mode additionally registers the \texttt{Virtual\allowbreak Thread\allowbreak Mount}/\texttt{Unmount} extension events and, as Elastic does, writes the mounted virtual thread's buffer address into a carrier-thread TLS slot on every transition~\cite{consumer}. The agent shares the buffer protocol with the JVM-allocation path and differs only in the notification and allocation channel, so the comparison isolates the capability tax defined in Section~\ref{JVMTI and Virtual Threads}.

\subsection{USDT Probes}
\label{USDT Probes}
To keep the modification small, we added the four probes to HotSpot's existing \texttt{hotspot.d} DTrace provider, so they behave like HotSpot's other probes; Appendix~\ref{app:probes} lists their definitions. The \texttt{freeze} and \texttt{thaw} probes fire in HotSpot's shared C++ continuation code (\texttt{continuationFreezeThaw.cpp}). The \texttt{start} and \texttt{end} probes fire from Java: on entering and leaving \texttt{run(task)}, \texttt{VirtualThread} calls a native method that fires the probe, once per virtual thread. These native methods are deliberately not intrinsic candidates, so the JIT cannot remove the calls (Section~\ref{Probe Placement: Constraints, Dead Ends, and Four-Probe Decomposition}). Their position also preserves the nested-lifetime rule for both allocation channels: \texttt{start} fires after the JVMTI start callback, so an agent-allocated address is already in place, and \texttt{end} fires before the JVMTI end callback, so the consumer drops its mapping before an agent frees the buffer.

\subsection{Gating}
\label{Gating}
Two switches, both fixed at JVM start-up, control the mechanism: the HotSpot product flag \texttt{-XX:+VThreadTraceProbes} enables the four USDT probes, and the system property \texttt{-Dvthread.\allowbreak trace.\allowbreak jvmAlloc=true} enables buffer allocation. Configurations S0--S2 in Table~\ref{tab:dormancy-gradient} are combinations of the two.

We first gated the probes with SDT semaphores, the usual USDT mechanism, but they cannot be used in Linux HotSpot. The build never defines the semaphore variables, so the JVM fails to link, and HotSpot's header include order records a semaphore address of zero in every probe note, so no tracer could enable them. HotSpot gates its own probes with product flags as well, and we follow this convention.

Like the JVMTI capability, the flag cannot be changed while the JVM is running; the difference is what it costs to leave on. With the flag on and no tracer attached, the probe sites are \texttt{NOP}s, and the cost (S2$-$S1) is below the measurement's resolution: at most about 33\,ns per event (Section~\ref{Observation overhead}). Enabling the JVMTI capability alone, by contrast, costs about 20\% on transition-heavy workloads (Section~\ref{Two channel comparison}). The flag can therefore stay on by default, and the cost of publication then depends only on whether a tracer attaches. This is the precise sense in which the cost of our mechanism is conditional on an observer being present.

\subsection{Consumers}
\label{Consumers}
To attribute an OS-level event, the consumer needs the event's carrier thread ID. As a representative OS-level event, we use the \texttt{tracepoint:\allowbreak syscalls:\allowbreak sys\_enter\_write} tracepoint: it fires on the OS side, independently of the JVM, and identifies only the carrier thread (\texttt{tid}), not the virtual thread that issued the call.

The consumer maintains a map from carrier thread ID to the mounted virtual thread's ID and buffer address, following the rules of Section~\ref{Consumer Model and Event-Level Equivalence}. Its correctness rests on two invariants: a carrier maps to at most one virtual thread at any time, which the insertion and deletion rules guarantee, and each mapping's lifetime lies within its buffer's lifetime, which the allocation points of Section~\ref{Buffer allocation} guarantee. When \texttt{sys\_enter\_write} fires, the consumer looks up the carrier's entry, reads the valid flag at the cached address, and either discards the record or reads the trace ID and span ID at their fixed offsets. It never writes to the buffer.

\subsection{OpenTelemetry SDK Integration}
\label{SDK Integration}
The buffer is written by the OpenTelemetry SDK through a \texttt{ContextStorage} decorator~\cite{consumer}. All reads and writes of the SDK's current context go through the \texttt{ContextStorage} SPI, so a decorator registered as an SDK plugin sees every change without any modification to OpenTelemetry itself; its only requirement is our patched JDK, whose \texttt{Thread.getTraceBufferAddress()} returns the buffer's address. The decorator follows one rule: after any change to the current context, such as attaching a span or closing a scope, the buffer reflects the new current context. Closing a scope restores the previous context, which the decorator republishes through the same path, so nested scopes need no special case: closing an inner scope republishes the outer span. If the restored context has no valid span, the decorator leaves the valid flag at 0. Each update follows the in-place rewrite protocol of Section~\ref{Trace Buffer Design}.

The publication path needs no changes to OpenTelemetry's source code, no bytecode weaving and no \texttt{-javaagent}: it depends only on a library-level SPI. The mechanism thus splits into three decoupled parts: the JDK (or a JVMTI agent) allocates the buffer and manages its lifetime, the USDT probes expose its address, and the SDK plugin writes trace context into it. The three communicate only through the \texttt{traceBufferAddress} field, which is why the two allocation channels of Section~\ref{Buffer allocation} are interchangeable: the writer cannot tell which one allocated the buffer.

\section{Evaluation}
\label{Evaluation}
We evaluate the mechanism against the three questions of Section~\ref{Motivation}: whether it attributes OS-level events correctly (Q1, Section~\ref{Correctness}), what it costs while no consumer is attached (Q2, Section~\ref{Observation overhead}), and what determines the cost of each publication channel (Q3, Sections~\ref{Observation overhead} and~\ref{Two channel comparison}).

\subsection{Setup}
\label{Overhead Methodology}
\paragraph{Configurations} We decompose the mechanism's cost into four configurations (Table~\ref{tab:dormancy-gradient}). We report differences between configurations rather than absolute times: each difference isolates one component, and the baseline cost common to both configurations cancels.

\begin{table}[t]
  \caption{Configurations used to decompose the mechanism's cost. Each adds one component to the previous one, so each difference between adjacent configurations isolates a single component. S0, the patched JDK with both switches off, is the reference: the Java-side gates are constants read once at \texttt{VirtualThread} class initialisation, the USDT sites are \texttt{NOP}s, and each HotSpot freeze/thaw site still tests the probe flag. S0 was not measured against an unpatched JDK.}
  \centering
  \small
  \setlength{\tabcolsep}{4pt}
  \label{tab:dormancy-gradient}
  \begin{tabular}{@{}llllll@{}}
    \toprule
    \textbf{State} & \textbf{Buffer} & \textbf{USDT} & \textbf{Consumer}
      & \textbf{Isolated by} & \textbf{Component} \\
    \midrule
    S0 & off & off & --      & --          & reference (+8\,B per vthread) \\
    S1 & on  & off & --      & S1$-$S0     & buffer allocation \\
    S2 & on  & on  & absent  & S2$-$S1     & publication residual \\
    S3 & on  & on  & attached & S3$-$S2    & observation \\
    \bottomrule
  \end{tabular}
\end{table}

\paragraph{Workloads} Three JMH microbenchmarks exercise different parts of the virtual-thread lifecycle:
\begin{itemize}
  \item \textbf{Transition}: one virtual thread calls \texttt{Thread.yield()} 10{,}000 times. Each yield is one freeze and one thaw, so the workload consists almost entirely of mount/unmount transitions.
  \item \textbf{Park/Unpark}: two virtual threads hand control to each other 10{,}000 times, each blocking in \texttt{LockSupport.\allowbreak park()} until the other calls \texttt{unpark()}; this hand-off is closer to real waits than \texttt{yield}.
  \item \textbf{Churn}: each operation starts 1{,}000 virtual threads with a near-empty body and joins them. The threads never park, so only \texttt{start} and \texttt{end} fire, and the workload isolates the per-thread lifecycle cost.
\end{itemize}

\paragraph{Measurement} We use JMH~1.37 with its default settings: average time per operation, five forks, and five 10\,s warmup and five 10\,s measurement iterations, so that JIT compilation settles during warmup and forks average out differences in code and heap layout between JVM instances. Performance results were collected on a bare-metal machine, and correctness runs on a virtual machine (Table~\ref{tab:Environment}).

\begin{table}[t]
  \centering
  \small
  \setlength{\tabcolsep}{4pt}
  \caption{Experimental environments.}
  \label{tab:Environment}
  \begin{tabular}{@{}lp{0.42\linewidth}p{0.34\linewidth}@{}}
    \toprule
    & \textbf{Performance (Sections~\ref{Observation overhead}--\ref{Two channel comparison})} & \textbf{Correctness (Section~\ref{Correctness})} \\
    \midrule
    Host        & AWS EC2 \texttt{c7i.metal-24xl}, bare metal & OrbStack VM, Apple-silicon host \\
    CPU         & x86-64, 96 logical CPUs (SMT 2), 1 NUMA node; \texttt{performance} governor, Turbo Boost disabled, no pinning & aarch64, 11 vCPUs \\
    Memory      & 188\,GiB & 15\,GiB \\
    OS / kernel & Ubuntu 24.04 LTS, Linux 6.17.0-1017-aws & Fedora 43, Linux 7.2.12 \\
    \texttt{bpftrace} & v0.20.2 & v0.24.2 \\
    \midrule
    Patched JDK & \multicolumn{2}{p{0.78\linewidth}}{OpenJDK 21.0.12-internal, tag \texttt{vthread-probes-v2}~\cite{jdkpatch}; SHA-256-verified release asset (x86-64), built from the same tag (aarch64)} \\
    Consumer    & \multicolumn{2}{p{0.78\linewidth}}{\texttt{ebpf-vthread-correlate} @ \texttt{18a6b80b}~\cite{consumer}} \\
    JMH         & 1.37, defaults: 5 forks, $5 \times 10$\,s warmup, $5 \times 10$\,s measurement, 1 thread & -- \\
    \bottomrule
  \end{tabular}
\end{table}

\paragraph{Uncertainty} Each JMH score is reported with JMH's default 99.9\% confidence interval half-width. For the difference between two configurations, we report the quadrature sum of the two half-widths, $\sqrt{e_1^2 + e_2^2}$, and mark a contrast \emph{n.s.} when the resulting interval contains zero. An n.s. contrast therefore means the effect is below the resolution of the measurement, not that it is zero; where it matters, we report the interval's upper bound.

\subsection{Correctness}
\label{Correctness}
We validate correctness (Q1) with an independent dual-channel harness, built on the definitions of Section~\ref{Correctness Definitions}. The ground truth is carried by a channel that never passes through the trace buffer of Section~\ref{Trace Buffer Design}: the harness determines each marker's expected trace context and virtual thread, while the eBPF consumer independently records the trace context and virtual thread it attributes to the same event. Each run uses 128 markers, from 16 virtual threads emitting 8 markers each. A marker is a single \texttt{write} system call issued by the virtual thread, whose length encodes the sender; the order of a sender's markers gives the sequence number~\cite{consumer}. The two scope modes in Table~\ref{tab:Correctness Validation Results} differ in whether the harness closes the final scope: \emph{leaky-final} deliberately violates scope discipline and leaves a valid record in the buffer when the virtual thread terminates. Each marker is classified independently on the content plane and the identity plane. The runs were performed on a local aarch64 virtual machine (Table~\ref{tab:Environment}).

\begin{table}[t]
  \centering
  \small
  \setlength{\tabcolsep}{3.5pt}
  \caption{Correctness validation results, summed over all runs of each configuration (128 markers per run). Identity-plane \emph{Unmapped} markers have no virtual thread in the consumer's map.}
  \label{tab:Correctness Validation Results}
  \begin{tabular}{@{}llrrrrrrrr@{}}
    \toprule
    & & & \multicolumn{4}{c}{Content plane} & \multicolumn{3}{c}{Identity plane} \\
    \cmidrule(lr){4-7} \cmidrule(l){8-10}
    Protocol & Scope mode & Runs & Correct & Unmapped & Invalid & Misattr. & Correct & Unmapped & Misattr. \\
    \midrule
    Four-probe & clean        & 5 & 640 & 0  & 0  & 0  & 640 & 0  & 0  \\
               & leaky-final  & 1 & 128 & 0  & 0  & 0  & 128 & 0  & 0  \\
    Two-probe  & clean        & 5 & 565 & 18 & 57 & 0  & 560 & 18 & 62 \\
               & leaky-final  & 1 & 114 & 4  & 0  & 10 & 112 & 4  & 12 \\
    \bottomrule
  \end{tabular}
\end{table}

The four-probe protocol attributed every marker correctly on both planes in all six runs (768/768), satisfying the criteria of Section~\ref{Correctness Definitions}. The result is the same under leaky-final, which shows that its correctness is ensured by the protocol rather than by the SDK's scope discipline.

The two-probe protocol failed as predicted in Section~\ref{Soundness and Completeness Analysis}, and it failed deterministically: every one of its six runs has exactly 16 failing markers, and they are always the first marker of each of the 16 virtual threads, emitted before the thread's first park. A virtual thread's first mount fires no \texttt{thaw}, so the two-probe consumer has no entry for it, and a terminated virtual thread never fires \texttt{freeze}, so its entry is never deleted. The first marker therefore finds either an empty slot (\emph{Unmapped}, 2--7 per run depending on scheduling) or a stale entry left by an earlier virtual thread on the same carrier. Only the split between these two outcomes varies between runs; the 16 failures do not.

What a stale entry produces depends on the buffer it points to, and this is where the two planes diverge. Under clean scopes, the terminated thread's buffer was left with \texttt{valid = 0}, so the content plane reports an honest \emph{Invalid}, while the identity plane still attributes the marker to the wrong virtual thread. Under leaky-final, the stale buffer was freed while still holding \texttt{valid = 1}: all 10 content misattributions passed the valid-flag check and returned a well-formed trace context of another virtual thread, read from released memory. This is the silent failure of C3, and it shows that the scope mode of the producer decides whether the same structural defect surfaces as an honest failure or as silent misattribution. Finally, in five of the six runs, 1--2 markers per run (7 in total) have a \emph{correct} trace context but a \emph{wrong} virtual thread. This is consistent with the allocator reusing the freed buffer address for the current virtual thread, so the stale entry happens to point at the right record. A content-only check would count these markers as correct, which is why C2 requires both planes.

Churn's freeze/thaw event count is zero, so the two-probe protocol's coverage on this workload is 0\%: a structural blind spot, not a probabilistic miss.

Both protocols use the same consumer and the same buffer protocol and differ only in which lifecycle events publish the buffer address. Every two-probe failure has the same root cause: the consumer's mapping does not follow the buffer's actual lifetime, either because no entry exists yet at the first mount or because a stale entry outlives a terminated virtual thread. The four-probe protocol succeeds because \texttt{start}/\texttt{end} close exactly these gaps, reconstructing the mount/unmount semantics at both ends of the lifecycle.

\subsection{Cost Decomposition}
\label{Observation overhead}
Table~\ref{tab:dormancy-gradient-results} reports the differences between the configurations. The first three rows give the dormant cost, with no consumer attached.

\paragraph{Allocation (S1$-$S0)} Allocation is measurable only on Churn, at $+377 \pm 30\,\mu s$ per operation. This is expected: the buffer is allocated once per virtual thread, and Churn creates 1{,}000 virtual threads per operation, against one for Transition and two for Park/Unpark. Dividing by 1{,}000 gives $0.377\,\mu s$ per virtual thread. Projected onto Transition, the same per-thread cost is several hundred times smaller than the measurement's resolution ($\pm 225\,\mu s$). The allocation cost is therefore consistent across the three workloads: it is paid once per virtual thread and becomes visible only when creating virtual threads dominates the workload. On Churn, whose baseline cost is about $0.41\,\mu s$ per virtual thread, it nearly doubles the cost of creating one.

\paragraph{Publication residual (S2$-$S1)} The USDT residual is non-significant on all three workloads: with the probes enabled but no consumer attached, the extra work is evaluating the probe arguments and executing a \texttt{NOP}, with no system call and no allocation. The point estimates on Transition and Park/Unpark are positive (+2.91\% and +2.35\%), and the intervals bound the effect at about +6\%. Normalised by the number of probe events per operation (see below), this is 15\,ns (upper bound 33\,ns) per event on Transition and 10\,ns (upper bound 26\,ns) on Park/Unpark.

\paragraph{Total dormant cost (S2$-$S0)} The total dormant cost is the sum of these two components: +2.47\% (n.s.) on Transition and +0.57\% (n.s.) on Park/Unpark, where neither component is resolved, and +89.38\% on Churn, that is, $0.371\,\mu s$ per virtual thread, consistent with S1$-$S0 within its uncertainty and therefore dominated by allocation. The dormant cost of our mechanism is therefore not zero: the publication path is free to within the measurement's resolution, but the buffer allocation is paid for every virtual thread, whether or not a consumer is present.

\begin{table}[t]
  \centering
  \small
  \setlength{\tabcolsep}{3pt}
  \caption{Cost decomposition: differences between the configurations of Table~\ref{tab:dormancy-gradient}. Cells show the difference $\pm$ propagated error in $\mu$s/op and the relative change; the percentage denominators are S0, S1, S0, S2 and S0, respectively. n.s.\ marks an interval containing zero.}
  \label{tab:dormancy-gradient-results}
  \begin{tabular}{@{}lrrr@{}}
    \toprule
    Contrast & Transition & Park/unpark & Churn \\
    \midrule
    S1$-$S0: allocation alone & \begin{tabular}{@{}r@{}}$-45 \pm 225$\\-0.43\% (n.s.)\end{tabular} & \begin{tabular}{@{}r@{}}$-308 \pm 517$\\-1.74\% (n.s.)\end{tabular} & \begin{tabular}{@{}r@{}}$+377 \pm 30$\\+90.86\%\end{tabular} \\
    S2$-$S1: USDT residual, real buffer & \begin{tabular}{@{}r@{}}$+307 \pm 351$\\+2.91\% (n.s.)\end{tabular} & \begin{tabular}{@{}r@{}}$+409 \pm 600$\\+2.35\% (n.s.)\end{tabular} & \begin{tabular}{@{}r@{}}$-6 \pm 19$\\-0.78\% (n.s.)\end{tabular} \\
    S2$-$S0: total dormant deployed & \begin{tabular}{@{}r@{}}$+262 \pm 325$\\+2.47\% (n.s.)\end{tabular} & \begin{tabular}{@{}r@{}}$+101 \pm 527$\\+0.57\% (n.s.)\end{tabular} & \begin{tabular}{@{}r@{}}$+371 \pm 31$\\+89.38\%\end{tabular} \\
    S3$-$S2: observation cost & \begin{tabular}{@{}r@{}}$+12977 \pm 347$\\+119.62\%\end{tabular} & \begin{tabular}{@{}r@{}}$+27111 \pm 903$\\+152.24\%\end{tabular} & \begin{tabular}{@{}r@{}}$-34 \pm 16$\\-4.28\%\end{tabular} \\
    S3$-$S0: total cost, consumer attached & \begin{tabular}{@{}r@{}}$+13238 \pm 219$\\+125.05\%\end{tabular} & \begin{tabular}{@{}r@{}}$+27212 \pm 850$\\+153.68\%\end{tabular} & \begin{tabular}{@{}r@{}}$+337 \pm 29$\\+81.28\%\end{tabular} \\
    \bottomrule
  \end{tabular}
\end{table}

\paragraph{Observation cost (S3$-$S2)} With a consumer attached, the cost is 119.62\% on Transition and 152.24\% on Park/Unpark. The two workloads fire different numbers of probe events per operation, 20{,}021 and 39{,}364; divided by these counts, the cost is $648 \pm 17$\,ns and $689 \pm 23$\,ns per event. The two per-event costs are close, which is consistent with observation cost scaling with the number of probe events rather than the number of virtual threads; with two workloads, this is consistency, not proof of a scaling law. Churn does not fit the same model: it fires only 2{,}000 probe events per operation (\texttt{start} and \texttt{end} for each of 1{,}000 virtual threads), spread across carrier threads on different CPUs. Its observation cost is $-34 \pm 16\,\mu s$ ($-4.28\%$), a statistically significant \emph{decrease}. Attaching a consumer cannot plausibly make the workload faster, and we have no explanation for this result; we report it as an unexplained measurement artefact of a workload dominated by thread creation and joining, not as a real speed-up.

\subsection{Two-Channel Comparison}
\label{Two channel comparison}
This experiment delivers the same buffer address through different channels while keeping everything else fixed. All four variants run on the patched JDK with the same buffer protocol and with the USDT probes off; they differ only in which channel allocates the buffer and whether the JVMTI agent is loaded:
\begin{description}
  \item[Baseline] no agent and no JVM allocation;
  \item[JVM alloc] the JVM allocates the buffer (our channel);
  \item[Agent] the JVMTI agent allocates the buffer;
  \item[JVMTI publish] the agent also publishes the buffer address to carrier TLS on every mount and unmount, as Elastic does.
\end{description}

\begin{figure}[t]
    \centering
    \includegraphics[width=\linewidth]{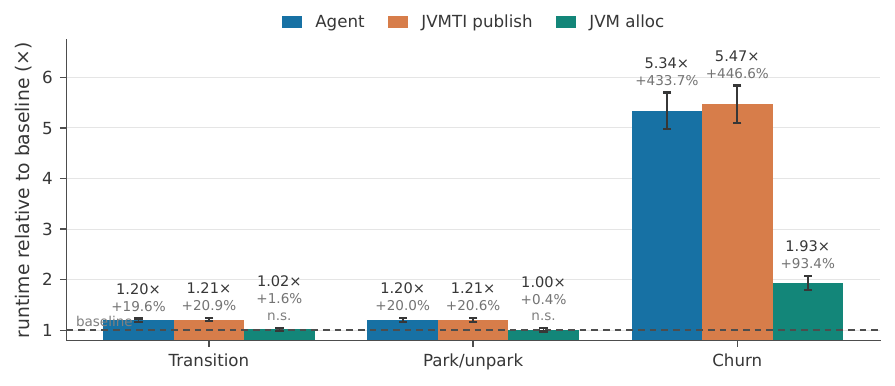}
    \caption{Overhead of the two publication channels relative to the baseline across the three workloads.
    \textnormal{n.s.} indicates that the corresponding uncertainty interval includes zero.}
    \Description{Grouped bar chart of runtime relative to the baseline for the Agent, JVMTI publish and JVM alloc variants on three workloads. On Transition and Park/Unpark, Agent and JVMTI publish are about 1.2 times the baseline and JVM alloc about 1.0 times. On Churn, Agent and JVMTI publish are about 5.3 to 5.5 times the baseline and JVM alloc about 1.9 times.}
    \label{fig:channel-overhead}
\end{figure}

Figure~\ref{fig:channel-overhead} (detailed data in Table~\ref{tab:main-matrix}) shows that JVM allocation is non-significant on Transition and Park/Unpark and costs $+93.4\%$ on Churn, that is, $0.38\,\mu s$ per virtual thread. This is the same component as S1$-$S0 in Table~\ref{tab:dormancy-gradient-results}, measured in a separate experiment, and the two agree.

The JVMTI channel costs about 20\% on Transition (+19.6\% Agent, +20.9\% JVMTI publish) and Park/Unpark (+20.0\%, +20.6\%). Two observations locate this cost. First, the Agent variant registers only \texttt{VirtualThreadStart}/\texttt{End}, which fire once per virtual thread, yet it already pays almost the full 20\% on workloads with one or two virtual threads per operation; adding per-transition TLS publication (JVMTI publish) adds only about one percentage point, within the measurement's uncertainty. The cost is therefore not the publication work itself, but the capability tax: loading the agent switches every mount and unmount onto the notifying transition path (Section~\ref{JVMTI and Virtual Threads}). Second, normalised by the same probe-event counts as above, the Agent cost is about 105\,ns per event on Transition and 89\,ns on Park/Unpark: similar across the two workloads, as expected of a per-transition tax. For comparison, the dormant USDT residual is bounded at 33\,ns per event.

On Churn, the JVMTI channel costs $+433.7\%$ (Agent) and $+446.6\%$ (JVMTI publish), that is, about $1.78$ and $1.83\,\mu s$ per virtual thread, roughly $4.6$--$4.8\times$ the $0.38\,\mu s$ of JVM allocation. Here the JVMTI path pays, for every virtual thread, two callbacks with JNI field accesses in addition to the allocation itself.

Both channels deliver the same 8-byte address, so the difference lies in the delivery channel, not in what is delivered. The JVMTI channel's cost is set by the capability tax and grows with the number of mount/unmount transitions, and it is paid whether or not an observer is attached. With no consumer attached, the JVM-allocation channel pays only at the two endpoints of each virtual thread's lifetime, so its cost grows with the number of virtual threads created.

\section{Discussion}
\label{Discussion}
\subsection{The Design Space, Not a Horse Race}
The two channels are different points in a design space rather than competitors. The JVMTI channel, which Elastic has shown to work, serves any reader: it writes the buffer address into ELF TLS, where every reader with access to the process's memory can find it, which is the generality OTEP 4947 aims for. Its price is the capability tax, paid on every transition for as long as the JVM runs, whether or not an observer is attached. Our USDT channel gives up that generality: a reader must be attached to the probes before the lifecycle events it needs fire, and must be able to read process memory, which in practice limits it to eBPF. In return, the cost of publication becomes conditional on a reader (Section~\ref{Gating}), and the only unconditional cost is the per-thread buffer allocation, which is significant only on creation-dominated workloads. On transition-heavy workloads, giving up reader-agnostic generality thus lowers the dormant price from around 20\% to below the measurement's resolution (Figure~\ref{fig:channel-overhead}). The design departs from the OTEP's TLS mechanism, but it is the kind of runtime-specific discovery mechanism the OTEP leaves room for, and it reuses the OTEP's record format.

Because observation cost follows probe events, and probe events follow mount and unmount transitions, an operator can estimate the cost of observing a workload from its transition rate: on our hardware, roughly 650--700\,ns per probe event (Section~\ref{Observation overhead}). Tasks that rarely unmount pay almost nothing, while tasks that block often pay in proportion to how often they do.

\subsection{Limitations}
Our evaluation has six limitations.
\begin{enumerate}
  \item \textbf{Workloads.} We use purpose-built microbenchmarks and harnesses, not application-level workloads such as Spring or Quarkus services.
  \item \textbf{Platforms and scale.} Our correctness evidence comes only from aarch64 and our performance evidence only from bare-metal x86-64. The correctness runs are also small (six per protocol, one for each leaky-final configuration), although the two-probe failure pattern was identical in every run.
  \item \textbf{Synthetic context.} The correctness harness writes synthetic trace contexts rather than real OpenTelemetry spans, so it validates the mapping protocol rather than the full OpenTelemetry stack. Conversely, because the verification path bypasses the SDK, the result does not depend on the SDK's scope discipline.
  \item \textbf{OBI.} We did not measure OBI's bytecode-instrumentation channel (Section~\ref{OBI}), which appeared concurrently with this work. Its per-transition system call places it with the JVMTI channel in our cost model, but we have not tested this.
  \item \textbf{Transition boundaries.} The four probes reconstruct the semantics of mount and unmount but not their precise boundaries (Section~\ref{Consumer Model and Event-Level Equivalence}); tools that need exact transition latencies would need true mount/unmount hooks.
  \item \textbf{Consumer cost.} Our benchmark consumer performs the full hot-path work of maintaining the map and counting events, but does not read the buffer. A production consumer adds a map lookup and a buffer read for every traced event, and the total depends on which events it traces.
\end{enumerate}

\section{Conclusion}
\label{Conclusion}
Under $M:N$ scheduling, the one-execution-per-thread assumption behind OS-level attribution no longer holds. We restored the join between OS-level events and units of work by binding a trace buffer to the lifetime of each virtual thread and publishing its address through four lifecycle probes, and we compared this channel with a JVMTI channel over the same buffer protocol. Our answers to the questions of Section~\ref{Motivation} are:
\begin{description}
  \item[Q1] Yes. The four-probe design satisfied the correctness criteria on both the content and the identity plane for all 768 test markers, while the two-probe design is unsound (Section~\ref{Correctness}).
  \item[Q2] Partly. The publication path is free to within the measurement's resolution while no consumer is attached: S2$-$S1 is non-significant on all three workloads, at most about 33\,ns per transition. The buffer allocation is not free: it costs $0.371\,\mu s$ per virtual thread, which is invisible on Transition and Park/Unpark but nearly doubles the cost of a creation-dominated workload (Section~\ref{Observation overhead}).
  \item[Q3] Each channel's cost is set by what it scales with. The JVMTI channel pays the capability tax on every mount and unmount: about 20\% on Transition and Park/Unpark, and about $1.8\,\mu s$ per virtual thread on Churn. The JVM-allocation channel pays only per virtual thread: non-significant on Transition and Park/Unpark, and $0.38\,\mu s$ per virtual thread ($+93.4\%$) on Churn (Section~\ref{Two channel comparison}).
\end{description}

Elastic and, concurrently, OpenTelemetry's eBPF instrumentation have achieved virtual-thread-level attribution by publishing on every transition. Our contributions are a different publication mechanism, whose dormant cost is a fixed allocation per virtual thread rather than a cost per transition, and which avoids the capability tax, and, to our knowledge, the first quantification of this cost structure in Java. Future work includes evaluating the mechanism on production web workloads, on both aarch64 and x86-64, and measuring OBI's bytecode-instrumentation channel within the same cost model.

\section*{Artifact and Data Availability}
The patched OpenJDK is available at~\cite{jdkpatch}, and the eBPF consumer, JVMTI agent, OpenTelemetry integration, benchmarks and correctness harness at~\cite{consumer}. The raw data behind every table and figure in Section~\ref{Evaluation}, namely the twelve correctness runs and the formal benchmark batch, is archived together with instructions for regenerating the tables and the figure~\cite{paperdata}.

\bibliographystyle{ACM-Reference-Format}
\bibliography{dissertation-arxiv}

\appendix

\section{Complete Benchmark Matrix}
\begin{table}[H]
  \centering
  \small
  \setlength{\tabcolsep}{3pt}
  \caption{Complete benchmark matrix. Time and allocation are raw JMH scores $\pm$ error; $\Delta t$ reports the difference from baseline and its relative change. n.s. denotes an interval containing zero. alloc. (B/op) reports managed-heap allocation only and excludes the 64\,B off-heap trace buffer.}

  \label{tab:main-matrix}
  \begin{tabular}{@{}llrrrr@{}}
    \toprule
    Workload & Variant & Time ($\mu$s/op) & $\Delta t$ ($\mu$s/op) & Relative & alloc. (B/op) \\
    \midrule
    Transition & Baseline & $10668.0 \pm 234.4$ & -- & -- & $240763.0 \pm 112.5$ \\
     & Agent & $12763.5 \pm 252.0$ & $2095.6 \pm 344.2$ & +19.6\% & $240844.9 \pm 138.2$ \\
     & JVMTI publish & $12897.8 \pm 204.9$ & $2229.8 \pm 311.3$ & +20.9\% & $240848.9 \pm 136.9$ \\
     & JVM alloc & $10838.9 \pm 167.7$ & $170.9 \pm 288.2$ & +1.6\% (n.s.) & $240763.7 \pm 114.3$ \\
    Park/unpark & Baseline & $17606.2 \pm 379.6$ & -- & -- & $460835.9 \pm 2151.7$ \\
     & Agent & $21125.5 \pm 478.3$ & $3519.3 \pm 610.7$ & +20.0\% & $465810.4 \pm 1392.0$ \\
     & JVMTI publish & $21228.4 \pm 454.7$ & $3622.2 \pm 592.4$ & +20.6\% & $464970.0 \pm 1063.6$ \\
     & JVM alloc & $17684.5 \pm 390.4$ & $78.2 \pm 544.6$ & +0.4\% (n.s.) & $456728.5 \pm 4045.9$ \\
    Churn & Baseline & $409.7 \pm 27.9$ & -- & -- & $320508.0 \pm 351.0$ \\
     & Agent & $2186.7 \pm 3.9$ & $1777.0 \pm 28.2$ & +433.7\% & $318446.6 \pm 25.5$ \\
     & JVMTI publish & $2239.6 \pm 7.3$ & $1829.8 \pm 28.9$ & +446.6\% & $318463.4 \pm 26.3$ \\
     & JVM alloc & $792.2 \pm 13.3$ & $382.5 \pm 30.9$ & +93.4\% & $316057.4 \pm 0.7$ \\
    \bottomrule
  \end{tabular}
\end{table}

\section{USDT Probe Definitions}
\label{app:probes}

\begin{lstlisting}[language=dtrace, caption={DTrace probes}]
probe vthread__freeze(uintptr_t, int, uintptr_t); //args: virtual thread id, freeze result, trace buffer address
probe vthread__thaw(uintptr_t, int, uintptr_t); //virtual thread id, thaw kind, trace buffer address
probe vthread__start(uintptr_t, uintptr_t); //virtual thread id,  trace buffer address
probe vthread__end(uintptr_t); //virtual thread id
\end{lstlisting}

\end{document}